\documentclass[envcountsame]{llncs}
\usepackage{amsfonts}
\usepackage{amsmath}
\usepackage{algorithm}
\usepackage{algorithmic}
\usepackage{graphicx,subfigure}

\renewcommand{\algorithmicrequire}{\textbf{Input:}}
\renewcommand{\algorithmicensure}{\textbf{Output:}}
\renewcommand{\algorithmiccomment}[1]{// #1}

\begin{document}

\title{Solving Medium-Density Subset Sum Problems in Expected Polynomial Time: An Enumeration Approach}

\author{Changlin WAN\inst{1,2}, Zhongzhi SHI\inst{1}}
\institute{Institute of Computing Technology, Chinese Academy of
Sciences\newline Beijing 100080, China \and Graduate University of
Chinese Academy of Sciences\newline Beijing 100080, China
\newline \email{changlin.wan@gmail.com}}  

\maketitle

\begin{abstract}
The subset sum problem (SSP) can be briefly stated as: given a
target integer $E$ and a set $A$ containing $n$ positive integer
$a_j$, find a subset of $A$ summing to $E$. The \textit{density} $d$
of an SSP instance is defined by the ratio of $n$ to $m$, where $m$
is the logarithm of the largest integer within $A$. Based on the
structural and statistical properties of subset sums, we present an
improved enumeration scheme for SSP, and implement it as a complete
and exact algorithm (EnumPlus). The algorithm always equivalently
reduces an instance to be low-density, and then solve it by
enumeration. Through this approach, we show the possibility to
design a sole algorithm that can efficiently solve arbitrary density
instance in a uniform way. Furthermore, our algorithm has
considerable performance advantage over previous algorithms.
Firstly, it extends the density scope, in which SSP can be solved in
expected polynomial time. Specifically, It solves SSP in expected
$O(n\log{n})$ time when density $d \geq c\cdot \sqrt{n}/\log{n}$,
while the previously best density scope is $d \geq c\cdot
n/(\log{n})^{2}$. In addition, the overall expected time and space
requirement in the average case are proven to be $O(n^5\log n)$ and
$O(n^5)$ respectively. Secondly, in the worst case, it slightly
improves the previously best time complexity of exact algorithms for
SSP. The worst-case time complexity of our algorithm is proved to be
$O(n\cdot 2^{n/2}-c\cdot 2^{n/2}+n)$, while the previously best
result is $O(n\cdot 2^{n/2})$.
\end{abstract}

\section{Introduction}

Let us denote $\mathbb{N}_+$ as the set of positive integers. The
subset sum problem is a classical NP-complete problem, in which one
asks, given a set $A = \left\{ {a_1 ,a_2 ,...,a_n } \right\}$ with
$a_j\in\mathbb{N}_+$ ($1\leq j \leq n$) and $E\in \mathbb{N}_+$, if
there exists a subset $A' \subseteq A$ such that the sum of all
elements of $A'$ is $E$. More formally, the subset sum problem can
be formulated as an integer programming problem:
\[
\begin{array}{l}
 \text{Maximize } z = \sum\limits_{j = 1}^n {a_j x_j }  \\
 \text{Subject to } \sum\limits_{j = 1}^n {a_j x_j  \le E} ;\forall j,x_j  = \text{0 or 1}. \\
 \end{array}
\]
Extensive study has been conducted on SSP and its related problems:
knapsack problem \cite{DBLP:books/kp/Martello90} and integer
partition problem \cite{DBLP:journals/rsa/Borgs01}. Many noticeable
results have been achieved. For example, the hardness distribution
of those problems are carefully investigated in
\cite{DBLP:journals/rsa/Borgs01}
\cite{DBLP:journals/jphy/Sasamoto01}
\cite{DBLP:journals/jsmte/Bauke04}
\cite{DBLP:journals/cor/Pisinger05} et al., and it is now known that
the hardness of SSP varies greatly with density $d$ (see
\cite{DBLP:conf/stac/Flaxman05}).

\textbf{Low-density:} an instance with density $0< d < c$, for some
constant $c$, can be efficiently solved by lattice reduction based
algorithms, e.g., \cite{DBLP:journals/jacm/Lagarias85}
\cite{DBLP:journals/siamjc/Frieze86}
\cite{DBLP:journals/cc/Coster92}. However, these algorithms have two
main limits. Firstly, they cannot solve instance with $d \geq c$
efficiently, though the bound of constant $c$ is recently extended
from 0.6463 to 0.9408. Secondly, they are not complete, i.e., they
may fail to find any solution of an instance when the instance
actually has solution.

\textbf{High-density:} an instance with density $d > c\cdot
n/\log{n}$ can be efficiently solved by various techniques such as
branch-and-bound, dynamic programming, and number theory analysis.
Specifically, the algorithm YS87 \cite{DBLP:journals/dam/Yanasse87}
adopts branch-and-bound technique; NU69
\cite{DBLP:journals/ms/Nemhauser69} and HS74
\cite{DBLP:journals/jacm/Horowitz74} adopt dynamic programming
technique; ST02 \cite{DBLP:journals/ejor/Soma02} adopts both
branch-and-bound and dynamic programming; CFG89
\cite{DBLP:journals/jc/Chaimovich89} and GM91
\cite{DBLP:/journals/siamjc/Galil91} utilize number theory analysis.
However, these algorithms have two main limits. Firstly, they cannot
solve instance with $d \leq c\cdot n/\log{n}$ efficiently. Secondly,
their average-case complexity is expected to increase with $n$, thus
they have difficulty in handling large size instance.

\textbf{Medium-density:} an instance with density $c \leq d$ and $d
\leq c\cdot n/\log{n}$ is usually hard to solve. As far as we know,
the algorithm DenseSSP \cite{DBLP:conf/stac/Flaxman05} is the only
previous algorithm that works efficiently in part of this density
scope. It solves uniformly random instances with density $d \geq
16n/(\log{n})^2$ in expected polynomial time $O(n^{3/2})$.

Other than exact algorithms, it is worth to mention that highly
efficient approximation methods (e.g.,
\cite{DBLP:journals/jacm/Ibarra75}
\cite{DBLP:journals/jcss/Kellerer03}) can solve SSP at polynomial
time and space cost. However, they cannot guarantee the exactness of
their solutions. In this paper, we concentrate on solving SSP
through exact methods, and we propose a complete and exact
algorithm, which we call EnumPlus. The two main ingredients of
EnumPlus are a new pruning mechanism and a new heuristic. Based on
the structural property of subset sums, the pruning mechanism allows
to dynamically partition the integer set into two parts and to prune
branches in the search tree. Based on the statistical property of
subset sums, the heuristic predicts which branch of the tree is more
likely to contain the solution (and this branch is explored first by
the algorithm).

\subsection{Contributions}\label{sec1.1}
The main contribution of this work is two-fold. First, by
equivalently reducing an instance to be low-density in linear time
(see Section \ref{sec4} and \ref{sec6.2}), we show the possibility
to design a sole algorithm that can efficiently solve arbitrary
density instance in a uniform way. Second, we propose a complete and
exact algorithm that has considerable advantage over previous exact
algorithms. Specifically, it extends the density scope, in which SSP
can be solved in expected polynomial time, and it slightly improves
the previously best worst-case time complexity of exact algorithms
for SSP.

\subsection{Notation and Conventions}\label{sec1.2}
If it is not specifically mentioned, we assume that the elements of
$A$ are sorted in decreasing order ($a_1>a_2>..>a_n$), and use $S$
to denote the sum of $A$. Following the notation and description
style of \cite{DBLP:books/kp/Martello90}, we denote some basic
notations that are used for the algorithm description as follows:
\begin{description}
  \item [$A_k$] denotes the subset $\left\{a_k, a_{k+1},..,a_n\right\}$ of $A$;
  \item [$S_k$] denotes the sum value of $A_k$ $\left( { = \sum\nolimits_{j = k}^n
{a_j } } \right)$;
  \item [$d_k$] denotes the density of $A_k$ $\left( { = \frac{n-k+1}{{\log max\{a_j|a_j\in
A_k\}}}} \right)$;
  \item[$W(E)$] denotes the number of solutions for a given target $E$ and
integer set $A$;
  \item [$\hat x_k$] denotes current partial solution $\left\{ {x_j =
0,1{\rm | }1 \le j \le k} \right\}$;
  \item[$\hat z_k$] denotes current partial solution value $\left( { =
\sum\nolimits_{j = 1}^k {a_j x_j } } \right)$;
  \item[$\hat c_k$] denotes current residual capacity $\left( { = E - \hat z_k } \right)$;
  \item[$\neg \hat c_k$] denotes current residual opposite capacity $\left( { =
S_{k+1} - \hat c_k } \right)$;
  \item[$b_{MAX} |\left( {A_k,\hat
c_k} \right)$] denotes the maximum subset sum of $A_k$ while
$b_{MAX}  \leq \hat c_k$;
  \item[$b_{MIN} |\left( {A_k,\hat c_k} \right)$] denotes the minimum subset sum of
$A_k$ while $b_{MIN} \geq \hat c_k$.
\end{description}

\section{Motivation}\label{sec2}
There are two main causes of performance discrepancy of different
enumeration (searching) scheme. In the first place, the efficiency
to prune infeasible solutions contributes to the performance both in
the worst case and in the average case. In the second place, proper
search strategy contributes to the performance in the average case.
Specifically, algorithm HS74 has the best time complexity $O(n\cdot
2^{n/2})$ in the worst case. It enumerates all possible solutions
following breadth first strategy; it prunes redundant branches by
dividing the original problem into two sub-problems and considering
all equal subset sums as one state. However, HS74 does not work well
in two situations. Firstly, when processing low-density instance,
almost all subset sums are different to each other, thus few pruning
can be made. Secondly, because of its breadth-first search strategy,
HS74 is slow to approach solutions when the size, i.e. breadth, of
an instance is considerable large.

The central idea of our approach is dynamically partitioning the
original instance $A[1..n]$ to two sub-instances $A[1..k]$ and
$A[k+1..n], 1<k<n$. We treat the whole enumeration space as a binary
tree (like the route colored by red in Figure \ref{fig1}) that is
stemmed from $A[1]$ and ended by $A[n]$. During the enumeration of
$A[1..n]$, all enumerated subset sums of $A[k+1..n]$ are organized
as ``block bounds", which serve as block barriers that can prevent
further expending of the $k$-th level nodes. Therefore, for any
partition point $k$, both $A[1..k]$ and A[k+1..n] are incrementally
and simultaneously enumerated by enumerating $A[1..n]$ as a binary
tree. In addition, a heuristic is utilized to accelerate the
searching for global solution. The heuristic predicts which branch
of the tree is more likely to contain the answer. Therefore, a large
problem is recursively reduced into a smaller one in linear time,
and it has high possibility that the two problems have at least one
common solution. To clarify the description of our algorithm, we
present the main phases separately.

\section{Branch and Prune}\label{sec3}
The pruning mechanism is inspired by the partition operation of
HS74. In HS74, the original instance is divided into two
sub-instances, and their subset sums are separately computed and
stored in two lists. For any subset sum $s_i$ in a list, if a subset
sum $s_j$ can be found in the other list such that $s_i+s_j=E$, a
feasible solution is located. While HS74 explicitly partitions the
the oriental instance only one time before enumeration, our
algorithm implicitly performs partition multiple times during
enumeration.

\begin{figure}[h]
    \centering
    \includegraphics[width=5.0in,bb=36 464 592 744]{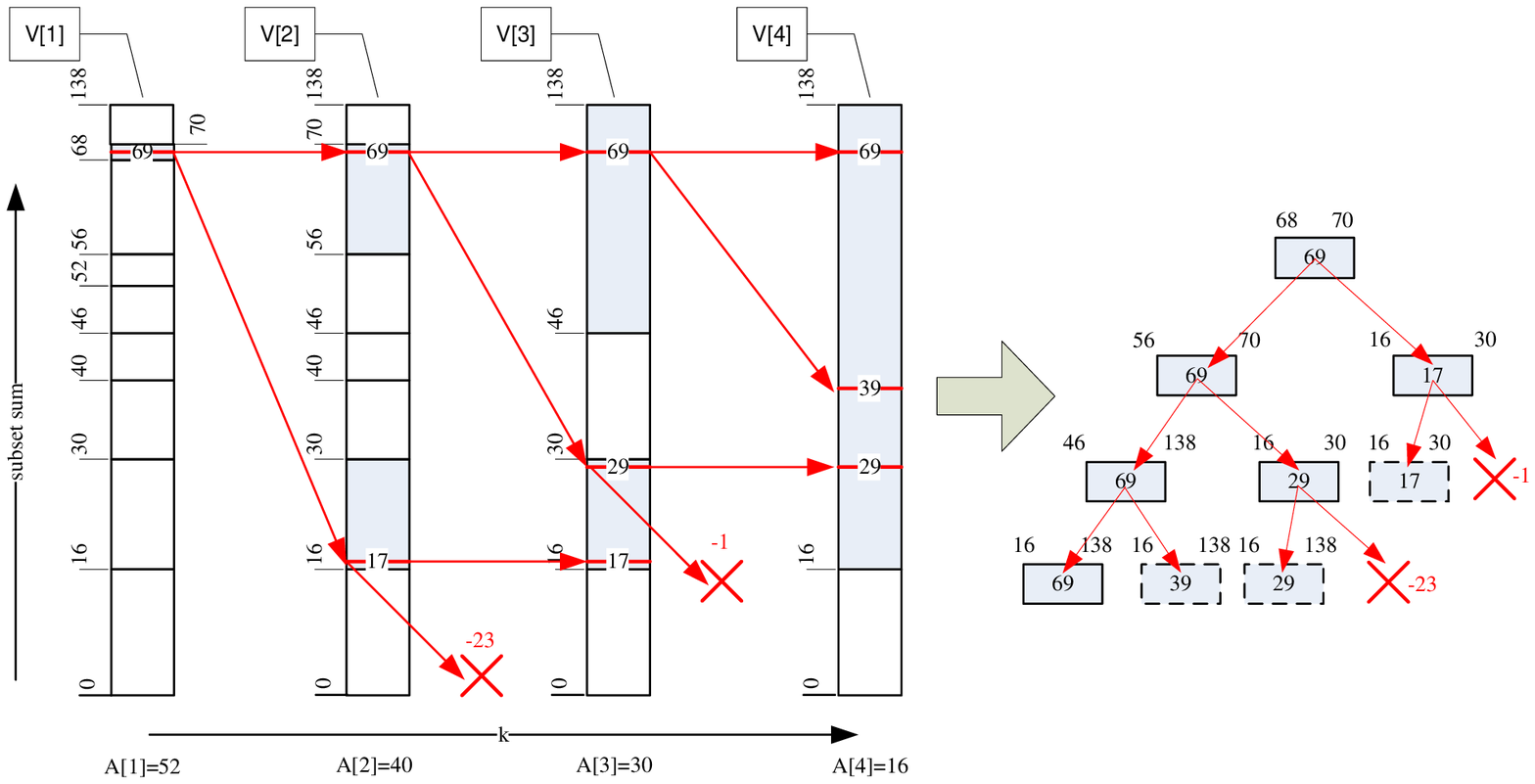}
    \caption{The generation of block bounds for target value $E=69$ and integer set $A[1..4] = \{52,40,30,16\}$.}
\label{fig1} 
\end{figure}

A ``block bound" of an integer set $A$ is defined as a two elements
structure $[b_{MAX}, b_{MIN}]$, in which $b_{MAX}$ and $b_{MIN}$ are
subset sums of $A$. Furthermore, a block bound must conform two
constraints: (1) $b_{MAX} < b_{MIN}$; (2) no subset sum of $A$ falls
between $b_{MAX}$ and $b_{MIN}$. Block bounds are recursively
calculated as follows:
\begin{displaymath}
\begin{array}{ll}
\text{If } \hat c_k \geq S_{k}, & b_{MIN} |(A_{k}, \hat c_k) = S, b_{MAX} |(A_{k}, \hat c_k) = S_{k}. \\
 \text{If } \hat c_k \leq 0, & b_{MIN} |(A_{k}, \hat c_k) = 0, b_{MAX} |(A_{k}, \hat c_k) = -
 a_{k+1}. \\
\text{If } S_{k} > \hat c_k > 0, & b_{MIN} |(A_{k}, \hat c_k) = min
\left\{
\begin{array}{l}
 b_{MIN} |(A_{k+1}, \hat c_k), \\
 a_{k} + b_{MIN}|(A_{k+1}, \hat c_k - a_{k}) \\
 \end{array} \right\},\\
 & b_{MAX} |(A_{k}, \hat c_k) =  max \left\{ \begin{array}{l}
b_{MAX} |(A_{k+1}, \hat c_k), \\
 a_{k} + b_{MAX} |(A_{k+1}, \hat c_k - a_{k}) \\
 \end{array} \right\}.\\
 \end{array}
\end{displaymath}
Let us consider a sorted integer array $A[1..n]$, we create an $n$
elements list $V[1..n]$. Each element $V[k]$ of $V[1..n]$ is a
collection of block bounds of the integer set $A_k$. Therefore, if
there is an integer $s = b_{MAX}$ (or $b_{MIN}$), $[b_{MAX},
b_{MIN}] \in V[k+1]$, and $s+\hat z_k = E$, a feasible solution for
target integer $E$ is located. If there is a block bound $[b_{MAX},
b_{MIN}] \in V[k+1]$ such that $b_{MAX} < \hat c_k < b_{MIN}$, we
can determine that there is no subset sum $s$ of $A_{k+1}$ such that
$s+\hat z_k = E$. In this way, a block bound $[b_{MAX}, b_{MIN}]$ of
$A_k$ acts as a bounded block that prevents all attempts to find
target $E$ in $A_k$ when $b_{MAX} \leq E \leq b_{MIN}$. To describe
the mechanism of block bound, a case that has an integer set
$A[1..4] = \{52,40,30,16\}$ and the target value $E = 69$ is
illustrated in Figure \ref{fig1}.

As we can see in Figure \ref{fig1}, the first node $\hat c_1=69$ is
expended to two nodes $\hat c_2=\{69, 17\}$, i.e., finding $E$ = 69
and 17 in subset $A[2..4]$. Suppose the node having larger $E$ is
always expended first, the first block bound $[16, 138]$ is
generated when finding $E = 69$ in subset $A[4..4]$. Therefore,
later finding of $E = 39, 29$ in subset $A[4..4]$ is blocked by the
block bound $[16, 138]$ . In the same way, let us observe the case
of $k=3$, the searching for $\hat c_3 = 29$ is finished with the
generation of a block bound $[16, 30]$, and the later searching for
$\hat c_3 = 17$ will be blocked by the block bound $[16, 30]$. When
the enumeration is finished, 6 block bounds $\{[68, 70], [16, 30],
[56, 70], [16, 30], [46, 138], [16, 138]\}$ are generated.

\section{Heuristic Search Strategy}\label{sec4}
Instead of pure depth-first or breadth-first search strategy, we
introduce a new heuristic to accelerate the approach to feasible
solution. At each state of enumeration, the expanding branch that
has larger possibility to find feasible solution will be explored
first. The heuristic is inspired by a previous study result of
\cite{DBLP:journals/jphy/Sasamoto01} in the context of canonical
ensemble, which is usually studied in the physics literature. The
main purpose of \cite{DBLP:journals/jphy/Sasamoto01} is to study the
property of the number of solutions in SSP, and then explain the
experiential asymptotic behavior of $W(E)$. As
\cite{DBLP:journals/jphy/Sasamoto01} suggested, given uniformly
random input integer set $A$ and target value $E$, the number of
solutions $W(E)$ is a central symmetric function with central point
at $E = S/2$. Moreover, $W(E)$ monotonically increases with the
increase of $E$ in $[0, S/2]$. If we denote ${\rm \textbf{Pr}}[E]$
as the possibility of that there exists at least one solution of
$E$, given two target value $E_1$ and $E_2$, we have that ${\rm
\textbf{Pr}}[E_1] > {\rm \textbf{Pr}}[E_2] {\text{ iff }}|S/2 - E_2
|
> |S/2 - E_1 |$. Suppose current partial solution is $\hat x_k$,
weather $x_{k+1} = 1$ (i.e. $\hat c_{k+1} = \hat c_k - a_{k+1}$) or
$x_{k+1} = 0$ (i.e. $\hat c_{k+1} = \hat c_k$) should be tried first
is decided by the inequation:
\begin{equation}
\label{eq1} \left| {{{\left( {\neg \hat c_k + \hat c_k} \right)}
\mathord{\left/
 {\vphantom {{\left( {\neg \hat c_k + \hat c_k} \right)} {2 - \left( {\hat c_k - a_{k+1} } \right)}}} \right.
 \kern-\nulldelimiterspace} {2 - \left( {\hat c_k - a_{k+1} } \right)}}} \right| > \left| {{{\left( {\neg \hat c_k + \hat c_k} \right)} \mathord{\left/
 {\vphantom {{\left( {\neg \hat c_k + \hat c_k} \right)} 2}} \right.
 \kern-\nulldelimiterspace} 2} - \hat c_k} \right|.
\end{equation}
Thus, we obtain the new heuristic: if inequation (\ref{eq1}) holds,
try $x_{k+1} = 0$ first, otherwise, try $x_{k+1} = 1$ first.

\section{The New Algorithm}\label{sec5}
Based on the ``block bound" and ``heuristic search" techniques, we
propose a complete and exact algorithm EnumPlus for SSP. In this
algorithm, the whole search space is enumerated as a binary tree
$T$. For any given target value $v$ at a branch node, the algorithm
try to find both $b_{MAX} |\left( {A_k,v} \right)$ and $b_{MIN}
|\left( {A_k,v} \right)$ in the sub-tree $T_k$ that has $x_k$ as
root node. If the block bound is already existed in the block bound
list $V[k]$, the existed block bound will be returned. Otherwise,
$T_k$ is expended to find the block bound $
[b_{MAX}|\left({A_k,v}\right),b_{MIN}|\left({A_k,v}\right)]$, and
the newly found block bound is inserted into $V[k]$ as a new
element. The enumeration procedure terminates in 2 cases: 1) a
feasible solution is found, 2) it is backtracked to the root of $T$.
At each branch node $(x_k)$ of $T$, if the target value $v$ is more
possible to be found when $x_k = 0$, then the branch $x_k = 0$ is
enumerated first, otherwise the branch $x_k = 1$ is enumerated
first.

The pseudo-code of EnumPlus and SetSum are given in Algorithm
\ref{alg1} and Algorithm \ref{alg2} respectively, while the concrete
implements of sub-algorithms QBB and UBB are not given since they
can be implemented by simply adopting some classic data
structures/algorithms (e.g., AVL-balance tree and Red-Black-balance
tree).
\begin{algorithm}
\caption{EnumPlus($A[1..N],E$)}\label{alg1}
\begin{algorithmic}[1]
  \REQUIRE an integer set $A[1..N]$; target value $E$.
  \ENSURE the maximum subset sum $b1 \leq E$; the minimum subset sum $b2 \geq
E$.
  \STATE allocate the vector of block bound sets $V[1..N]$;
  \STATE $S\Leftarrow$ sum value of $A[1..N]$;
  \STATE $[b1, b2]\Leftarrow {\rm SetSum}(1, E)$;
  \STATE destroy the vector of block bound sets $V[1..N]$;
  \RETURN $[b1, b2]$;
\end{algorithmic}
\end{algorithm}

\begin{algorithm}
\caption{SetSum($k, v$)} \label{alg2}
\begin{algorithmic}[1]
  \REQUIRE $k$ = start position of residual subset $A[k..n]$;

$v$ = residential capacity $\hat c$.

  \ENSURE $[b_{MAX}, b_{MIN}]$ = block bound of $A[k..N]$ for $\hat c = v$.

  \STATE \textbf{if} {$v \geq S_k$} \textbf{return} $[S_k, S]$;
  \STATE \textbf{if} {$v \leq 0$} \textbf{return} $[-a_{k+1}, 0]$;
  \STATE $[b_{MAX}, b_{MIN}] \Leftarrow {\rm QBB}(V[k], v)$;
  \COMMENT {query block bound $[b_{MAX}, b_{MIN}]$ in $V[k]$ such that $b_{MAX}\leq v \leq b_{MIN}$.}
  \IF {$b_{MAX}\leq v \leq b_{MIN}$}
    \IF {$v=b_{MAX}$ or $v=b_{MIN}$}
        \STATE identify solution; halt; \COMMENT
{found solution}
    \ELSE
        \RETURN $[b_{MAX}, b_{MIN}]$;
    \ENDIF
  \ELSIF {inequation \ref{eq1} holds}
      \STATE $[b3,b4] \Leftarrow {\rm SetSum}(k+1, v)$;
      \STATE $[b1,b2] \Leftarrow {\rm SetSum}(k+1, v - A[k])$; $b1 += A[k]; b2 += A[k]$;
  \ELSE
        \STATE $[b1,b2] \Leftarrow {\rm SetSum}(k+1, v - A[k], v2)$; $b1 += A[k]; b2 += A[k]$;
        \STATE $[b3,b4] \Leftarrow {\rm SetSum}(k + 1, v)$;
  \ENDIF
  \STATE $b_{MAX} \Leftarrow \max \{ b1,b3\}; b_{MIN} \Leftarrow \min \{b2,b4\}$;
  \STATE ${\rm UBB}(V[k], b_{MAX}, b_{MIN})$; \COMMENT{insert $[b_{MAX}, b_{MIN}]$ into $V[k]$}
  \RETURN $[b_{MAX}, b_{MIN}]$;
\end{algorithmic}
\end{algorithm}

\section{Performance Analysis}\label{sec6}
Before analyzing the complexity of our algorithm, we assume that the
requirement of time and space of our algorithm is maximized when
target value $E=S/2$. The assumption is reasonable because our
algorithm simultaneously search both $E$ and $S-E$ in the answer
space. Moreover, $E=S/2$ is the hardest case for the dynamic
programming algorithm (see \cite{DBLP:/journals/siamjc/Galil91}).
Therefore, all our following analysis will be provided in case of
that $S/2$ is chosen as target value $E$.

\subsection{Worst-Case Complexity}\label{sec6.1}
Before the presentation of our results about the worst-case
complexity of EnumPlus, we first introduce a lemma as follows:
\begin{lemma}\label{lem1}
For a certain subset $A[k..n]$ of $A$, the number of block bounds
generated by SetSum is less than $min\{2^{k-1},2^{n-k+1}\}$.
\end{lemma}
\begin{proof}
In case of $2^{k - 1}  \ge 2^{n - k + 1}$, the number of all
possible subset sums of $A[k..n]$ is less than $\left( {2^{n - k +
1} - 1} \right)$, therefore the number of all possible block bounds
of $A[k..n]$ is less than $2^{n - k + 1}$, i.e.
$min\{2^{k-1},2^{n-k+1}\}$. In case of $2^{k - 1} < 2^{n - k + 1}$,
the search tree has at most $2^{k - 1}$ nodes at level $k$. Because
each node generates at most one block bound, the number of all
possible block bounds of $A[k..n]$ is less than $2^{k - 1}$, i.e.
$min\{2^{k-1},2^{n-k+1}\}$.
\end{proof}

About the worst-case complexity of EnumPlus, there are 2
propositions given as follows:
\begin{proposition}\label{prop1}
The worst-case space complexity of EnumPlus is $O(2^{n/2})$.
\end{proposition}
\begin{proof}
For a subset $A_k$, the number of block bounds generated by SetSum
is less than $\min \{ 2^{k - 1} ,2^{n - k + 1} \}$, therefore the
total number ${\rm Num}(n)$ of generated block bounds is
\[
{\rm Num}(n) \leq \sum\limits_{k = 1}^n {\min \{ 2^{k - 1} ,2^{n - k
+ 1} \}} \le 2 \times \sum\limits_{k = 1}^{n/2} {2^{k - 1} } \le
2^{n/2 + 1}.
\]
Thus the worst-case space complexity of EnumPlus is $O(2^{n/2})$.
\end{proof}

\begin{proposition}\label{prop2}
The worst-case time complexity of EnumPlus is $O(n\cdot
2^{n/2}-c\cdot 2^{n/2}+n)$.
\end{proposition}
\begin{proof}
As we proved in the proposition \ref{prop1}, there are at most
$2^{n/2+1}$ block bounds are generated, and each recursive call for
SetSum generates one block bound. The main time cost of each block
bound is to search and insert it in a collection $V[k]$. There are
some classic data structures/algorithms, such as AVL-balance tree
and Red-Black-balance tree, can efficiently manage the search and
insert operations on storable data collection. The worst-case time
cost of these algorithms to search or insert in the $n$ elements
collection is $\log n$. Therefore we have the worst-case time cost
${\rm Time}(n)$ of EnumPlus as follows:
\begin{displaymath}
\begin{array}{ccll}
  {\rm Time}(n) & = & 2 \times \sum\limits_{k = 1}^{n/2} {\sum\limits_{i = 1}^{2^{k - 1} } {\left\lceil {\log i} \right\rceil }}  = 2 \times \sum\limits_{k = 1}^{n/2}
{\sum\limits_{i = 1}^k {((i-1) \times 2^{i - 2} )} } \\
   & = & 2 \times
\sum\limits_{k = 1}^{n/2} {((k - 2) \times 2^{k-1} + 1}) \\
 & \leq & (n-6)\times 2^{n/2} + n + 8
\end{array}
\end{displaymath}
Thus the worst-case time complexity of algorithm EnumPlus is
$O(n\cdot 2^{n/2}-c\cdot 2^{n/2}+n)$.
\end{proof}

\subsection{Average-Case Complexity}\label{sec6.2}
Before analyzing the average-case complexity of our algorithm, 2
lemmas are introduced as follows:
\begin{lemma}\label{lem2}
EnumPlus always reduces an instance $A_1$ with $\hat c_1 = S_1/2$ to
$A_k$ with $\hat c_k$, $|\hat c_k - S_{k}/2| \leq a_{k-1}/2$, in
linear time.
\end{lemma}
\begin{proof}
\textbf{[Induction]} We first consider $k=2$. Because of the
heuristic search strategy, EnumPlus first expends the branch that
leads to a sub-problem, in which $\left| {\hat c_k - S_k/2} \right|$
is smaller. Then we have
\[
\hat c_2  = \left\{ \begin{array}{l}
 S_1/2,{\rm  if }\left| {S_1 - S_2 } \right| \le \left| {S_1 - 2a_1 - S_2 } \right| \\
 S_1/2 - a_1,{\rm  if }\left| {S_1 - S_2 } \right| > \left| {S_1 - 2a_1 - S_2 } \right|. \\
 \end{array} \right.
\]
Therefore,
\[
\hat c_2 - S_{2}/2 = \left\{ \begin{array}{l}
 a_1/2,{\rm  if }\left| {S_1 - S_2 } \right| \le \left| {S_1 - 2a_1 - S_2 } \right| \\
 -a_1/2,{\rm  if }\left| {S_1 - S_2 } \right| > \left| {S_1 - 2a_1 - S_2 } \right|. \\
 \end{array} \right.
\]
Thus EnumPlus reduces $A_1$ with $\hat c_1 = S_1/2$ to $A_2$ with
$\hat c_2$, $|\hat c_2 - S_{2}/2| \leq a_{1}/2$, in 1 step.

Then we assume that EnumPlus reduces $A_1$ with $\hat c_1 = S_1/2$
to $A_k$ with $\hat c_k$, $|\hat c_k - S_{k}/2| \leq a_{k-1}/2$, in
$k-1$ steps.

Consider $A_{k+1}$ and $\hat c_{k+1}$, we have
\[
\hat c_{k+1}  = \left\{ \begin{array}{l}
 \hat c_{k},{\rm  if }\left| {\hat c_{k} - S_{k+1} } \right| \le \left| {\hat c_{k} - 2a_{k+1} - S_{k+1} } \right| \\
 \hat c_{k} - a_{k},{\rm  if }\left| {\hat c_{k} - S_{k+1} } \right| > \left| {\hat c_{k} - 2a_{k+1} - S_{k+1} } \right|. \\
 \end{array} \right.
\]
Combine the above definition of $\hat c_{k+1}$ and the assumption
that $|\hat c_k - S_{k}/2| \leq a_{k-1}/2$, we have that $|\hat
c_{k+1} - S_{k+1}/2| \leq a_{k}/2$. Thus EnumPlus reduces $A_1$ with
$\hat c_1 = S_1/2$ to $A_{k+1}$ with $\hat c_{k+1}$, $|\hat c_{k+1}
- S_{k+1}/2| \leq a_{k}/2$, in $k$ steps.
\end{proof}

\begin{lemma}\label{lem3}
Let $M=2^m$ and the $n$ elements of $A$ is uniformly random in
$[1..M]$, the number of distinct subset sums of $A$ is expected to
be $O(n^4)$.
\end{lemma}
\begin{proof}
We use $S_1,...,S_M$ to denote the sequence of all subsets of $A$
listed in non-decreasing order of their sums. Let the sum of subset
$S_u$ be $P_u=\sum\nolimits_{j \in S_u } {a_j }$. For any $2\leq u
\leq M$, define $\Delta_u=P_u-P_{u-1}\geq 0$, then $P_u$ is a
distinct subset sum if $\Delta_u > 0$, and $P_1$ is always a
distinct subset sum. Let every element $a_j$ of $A$ be a
non-negative random variable with density function
$f_j:[1..M]\rightarrow [0,1]$, i.e., $f_j(t)={\rm
\textbf{Pr}}(a_j=t), t\in [1..M]$. We notice that there is a
theorem, which is proved by \cite{DBLP:conf/stoc/BeierV03} for
general discrete distributions, shows that:

\textit{Suppose} $\pi = max_{j\in [1..n]}(max_{x\in [1..M]}(f_j(x))$
\textit{and} $\mu \geq max_{j\in [1..n]}({\rm \textbf{E}}[a_j])$.
\textit{Then the expected number of dominating sets is} ${\rm
\textbf{E}}[q] = O(\mu n^2(1-e^{-\pi n^2})) = O(\mu\pi n^4)$.

Because a distinct subset sum is a special case of dominating set on
condition that weight $w_j$ and profit $p_j$ are both equal to
$a_j$, the number of distinct subset sums is equal to the number of
dominating sets on this condition. Since $a_j$ is uniformly random
in $[1..M]$, we have $\pi = 1/M$, $\mu = M/2$. Therefore, the number
of distinct subset sums is expected to be
\[
{\rm \textbf{E}}[q] = O(\frac{M}{2} \cdot \frac{1}{M} \cdot n^4) =
O(n^4).
\]
\end{proof}

Assume that the elements of $A$ is uniformly random in $[1..M]$, we
have 2 propositions about the complexity of EnumPlus in the average
case:

\begin{proposition}\label{prop3}
Given an integer set $A$ whose elements are uniformly distributed,
the overall expected time and space requirement of EnumPlus in the
average case are $O(n^5\log n)$ and $O(n^5)$.
\end{proposition}
\begin{proof}
According to Lemma \ref{lem3}, the number of distinct subset sums of
$A$ is expected to be $O(n^4)$. Therefore the expected space cost is
$O(n \cdot n^4) = O(n^5)$, and the expected time cost is $O(n \cdot
n^4 \cdot \log {(n^4)}) = O(n^5\log n)$. Thus the overall time and
space complexity of EnumPlus are expected to be $O(n^5\log n)$ and
$O(n^5)$ respectively.
\end{proof}

\begin{proposition}\label{prop5}
EnumPlus solves SSP in $O(n\log n)$ time when density $d \geq c\cdot
\sqrt{n}/\log{n}$.
\end{proposition}
\begin{proof}
Consider an integer set $A[1..n]$ whose elements are uniformly
random in $[1..2^m]$. As the previous result shown by
\cite{DBLP:journals/jphy/Sasamoto01} and
\cite{DBLP:journals/rsa/Borgs01}, if density $d > 1$, there is a
high possibility that the instance with target value $S/2$ has many
solutions. Thus it is expected that the sub-instance $A_{n-m+1}$
with $S_{k-m+1}/2$ has many solutions, and it takes at most
$O(m2^{m/2})$ time to locate these solutions. Furthermore, as we
proved in Lemma \ref{lem2}, EnumPlus reduces instance $A[1..n]$ with
$S/2$ to sub-instance $A_k$ with $S_k/2$ in linear time. Thus the
expected time complexity for the problem $A$ with $S/2$ is
$O(m2^{m/2}) + O(n)$. If $n=O(2^{m/2l})$, $\frac{m}{2\log m} > l
\geq 1$, then the expected time complexity for instance $A[1..n]$
with $S/2$ is $O(n^l\log n)$, and $d=O(\sqrt[2l]{2^m}/m) =
O(\sqrt[2l]{n}/\log n)$. Thus EnumPlus solves SSP in $O(n\log n)$
time when density $d \geq c\cdot \sqrt{n}/\log{n}$.
\end{proof}

\subsection{Comparison of Related Works}\label{sec6.3}
Among the previously exact algorithms for SSP, HS74 has the best
time complexity $O(n\cdot 2^{n/2})$ in the worst case. EnumPlus is
an overall improvement of HS74, its worst-case time complexity is
$O(n\cdot 2^{n/2}-c\cdot 2^{n/2}+n)$. As we described in Section
\ref{sec2}, HS74 always reduce the original instance to 2 half size
sub-instances. As we know, if the density of sub-instance is larger
than 1, a solution is expected to be found by solving one
sub-instance whose size is half of the original instance. However,
by using a new heuristic, EnumPlus reduce the original instance to a
smaller sub-instance, in which the solution can be found (see the
proof of Proposition \ref{prop5}). Thus the performance of EnumPlus
is better than HS74 in average case, especially when handling large
size instance. Specifically, EnumPlus solves SSP in $O(n\log n)$
time when density $d \geq c\cdot \sqrt{n}/\log{n}$. This density
bound is better than the density bound $d \geq c\cdot n/(\log{n})^2$
of DenseSSP, which is the only previous algorithm working
efficiently beyond the magnitude bound of $O(n/\log n)$. However, it
must be noticed that the performance of EnumPlus is still not good
enough when handling low-density instance. When density $d <
0.9408$, some incomplete algorithms, which are based on lattice
reduction, are expected to outperform EnumPlus.

\section{Conclusions and Future Work}\label{sec7}
In this work, we proposed a new enumeration scheme that utilizes
both structural property and statistical property of subset sums to
improve the efficiency of enumeration. The improved enumeration
scheme is implemented as a complete and exact algorithm (EnumPlus).
The algorithm always equivalently reduces an instance to be
low-density, and then solve it by enumeration. Through this
approach, we show the possibility to design a sole algorithm that
can efficiently solve arbitrary density instance in a uniform way.
Furthermore, our algorithm has considerable performance advantage
over previous exact algorithms. It slightly improves the previously
best time complexity of exact algorithms for SSP in the worst case;
it extends the density scope to $d \geq c\cdot \sqrt{n}/\log{n}$, in
which SSP can be solved in polynomial time. In addition, the overall
expected time and space requirements are proved to be $O(n^5\log n)$
and $O(n^5)$ respectively in the average case.

As we previously described, arbitrary density SSP instance can be
equivalently reduced to and solved as low density instance by our
approach. Thus the efficiency of EnumPlus mainly relies on
efficiently solving low density problem. Since the lattice reduction
approach shows particular efficiency when dealing low density
instance, the integration of the two approaches may be a potential
way to further improve the performance of our algorithm. Therefore,
the relationship between lattice reduction and enumeration scheme is
an important issue in our future work.

\bibliographystyle{splncs}
\bibliography{EnumPlus}

\end{document}